# Microcavity-enhanced Kerr nonlinearity in a vertical-external-cavity surface-emitting laser

C. Kriso[1], S. Kress[1], T. Munshi[1], M. Großmann[2], R. Bek[2], M. Jetter[2], P. Michler[2], W. Stolz[1], M. Koch[1], and A. Rahimi-Iman[1*]

[1]*Faculty of Physics and Materials Sciences Center, Philipps-Universität Marburg, D-35032 Marburg, Germany*

[2]*Institut für Halbleiteroptik und Funktionelle Grenzflächen, Universität Stuttgart, D-70569 Stuttgart, Germany*

**Self-mode-locking has become an emerging path to the generation of ultrashort pulses with vertical-external-cavity surface-emitting lasers. In our work, a strong Kerr nonlinearity that is so far assumed to give rise to mode-locked operation is evidenced and a strong nonlinearity enhancement by the microcavity is revealed. We present wavelength-dependent measurements of the nonlinear absorption and nonlinear-refractive-index change in a gain chip using the Z-scan technique. We report negative nonlinear refraction up to $1.5 \cdot 10^{-11}$ cm$^2$/W in magnitude in the (InGa)As/Ga(AsP) material system close to the laser design wavelength, which can lead to Kerr lensing. We show that by changing the angle of incidence of the probe beam with respect to the gain chip, the Kerr nonlinearity can be wavelength-tuned, shifting with the microcavity resonance. Such findings may ultimately lead to novel concepts with regard to tailored self-mode-locking behavior achievable by peculiar Kerr-lens chip designs for cost-effective, robust and compact fs-pulsed semiconductor lasers.**

The Kerr effect is at the basis of many important device concepts like all-optical switching[1], optical limiting[2] and soliton mode-locking of lasers and microresonators[3,4]. The capability to accurately measure and model the nonlinear refractive index changes associated with the Kerr effect is crucial for improved device operation, where a specifically tailored nonlinear refractive index is required, e.g. for the intensity-dependent Kerr lensing. Several measurement schemes have been developed in the past for the characterization of the nonlinear refractive index[5–7], with the Z-scan technique[8] being undoubtedly the most prominent one due to its simplicity and high sensitivity. This method continues to be of high experimental value with the rise of novel material classes like graphene and other two-dimensional semiconductors which often exhibit a very strong refractive nonlinearity[9–11].

Kerr-lens mode-locked Ti:Sapphire lasers have dominated the field of ultra-short high-power mode-locked lasers since their initial discovery nearly three decades ago[12,13]. In these lasers, the intensity-dependent refractive index of the gain crystal leads to self-focusing of the laser beam for high intensities. When part of the continuous-wave (cw) beam profile is suppressed by inserting a slit into the cavity or reducing the pump spot on the crystal, an artifical ultrafast saturable absorber can be formed and can lead to very short pulse emission down to a few fs in duration[14,15].

In recent years, a similar behavior has been observed in saturable-absorber-free mode-locked vertical-external-cavity surface-emitting lasers (VECSELs)[16–20] with very good mode-locking properties being reported for quantum-well-based (QW)[21] as well as for quantum-dot-based VECSELs[22]. These "self-mode-locked" VECSELs could lead to cheap and compact, high-power pulsed sources with sub-GHz to multi-GHz repetition rates for frequency metrology, spectroscopy and nonlinear imaging, rendering expenses for and limitations of saturable-absorber mirrors obsolete. Thus, one can expect self-mode-locking to open up new application scenarios for these techniques. However, the primary question which arose in this context yet remains unanswered, namely which effects are governing and promoting stable mode-locked operation with ultrafast sub-ps pulse emission.

Here, we present wavelength-dependent measurements of the nonlinear refractive index around the resonance of a multi-quantum-well VECSEL gain structure designed for lasing at around 960 nm. We perform Z-scan measurements in the range of 930 nm to 975 nm in order to reveal the interplay of nonlinear refraction and absorption in this spectral range. Changing the angle of incidence allows us to discuss the role of the longitudinal confinement factor, i.e. the cavity resonance, in the observed enhancement of the nonlinearity. We report a negative nonlinear refraction up to $1.5 \cdot 10^{-11}$ cm$^2$/W in magnitude close to the laser design wavelength. This leads to a defocusing lens which is sufficient to perturb the cavity beam and give rise to Kerr-lens mode-locking. ***Figure 1a*** displays a possible cavity configuration to exploit this measured Kerr-lens for mode-locking where the defocusing lens in the gain chip leads to beam narrowing at the end mirror, thus, favoring mode-locking when a slit is inserted there. A similar cavity has been used in Ref. [21] to obtain self-mode-locking and has been investigated numerically in Ref. [23] to support in principle Kerr-lens mode-locking for a defocusing nonlinear lens in the gain chip.

**1. Investigations on the self-mode-locking effect in VECSELs**

The claim of the observed „self-mode-locking" behaviour being governed by Kerr-lens mode-locking[17] has been supported by investigations of the nonlinear refractive index in VECSEL gain structures[24–26], reporting refractive index changes sufficiently strong to possibly perturb the cavity beam profile in a way to cause mode-locking. However, these measurements were exclusively performed at arbitrarily selected (experimentally available) single wavelengths. Such experiments were not taking into account the microcavity resonance as well as the strong dispersion of the nonlinear refractive index around the band edge, which is characteristic for contributions to the nonlinear refractive index, both, from the bound-electronic Kerr effect (BEKE) existing below the bandgap[27] and from free-carrier-related nonlinearities (FCN) existing above the band edge[28].

While the nonlinear optical properties of semiconductors below the band gap have been extensively studied and modeled using the Z-scan method[27,29,30], nonlinear optical processes above and around the band gap have mostly been investigated by pump-probe techniques [28,31] or linear absorption measurements [32]. These methods cannot access the nonlinear refractive index directly but use nonlinear Kramers-Kronig relations to calculate it from nonlinear absorption spectra. Very few reports on direct measurements of the nonlinear refractive index properties of QW structures exist so far which solely can account fully for nonlinear pulse propagation of a single pulse through the medium[33,34].

Although giving a rough estimate of the strength of the nonlinearity in multiple-quantum-well structures, these measurement results cannot arbitrarily be transferred to other sample designs as the nonlinear refractive index constitutes an effective one, composed of different material nonlinearites and their respective interaction containing both BEKE and FCN. Thus, the direct determination of the nonlinear refractive index by Z-scan measurements becomes inevitable whenever its precise value for a particular semiconductor heterostructure is of interest. We define the effective nonlinear refractive index $n_2$ as

$$n(I) = n_0 + n_2 I, \qquad (1a)$$

with $n(I)$ being the total refractive index, $n_0$ the intensity independent refractive index and $I$ the optical peak power density. Complementary to the definition of the intensity dependent refractive index $n_2$ the nonlinear absorption $\beta$ can be defined as

$$\alpha(I) = \alpha_0 + \beta I, \qquad (1b)$$

with $\alpha_0$ being the linear absorption coefficient and $\alpha(I)$ the total absorption coefficient. It is important to point out that $\beta$ compared to $n_2$ only comprises ultrafast $\chi^{(3)}$ effects, as carrier-related absorption would lead to saturation and thus more complex relation on incident intensity as detailed in the Supplementary Information.

The goal of this work is the determination of this effective $n_2$ of a VECSEL chip as well as of the nonlinear absorption $\beta$ acting as possible loss mechanism.

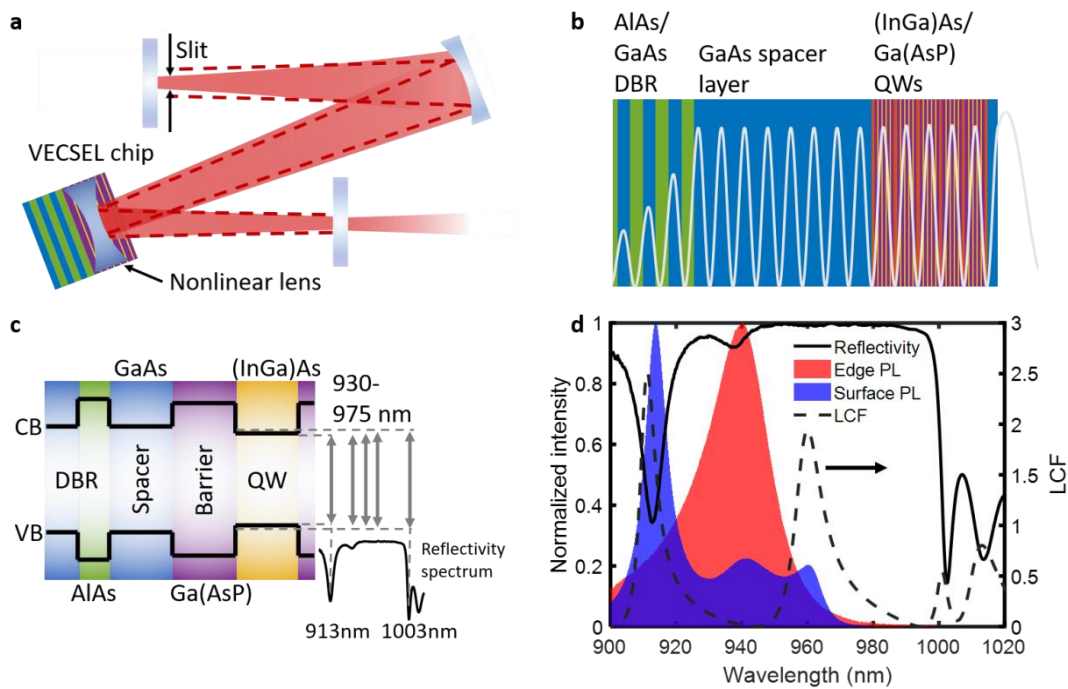

**Figure 1 | VECSEL-chip characteristics:** (a) Illustration of a Kerr-lens mode-locked VECSEL with the dashed lines indicating the cw beam profile while the shaded area traces the beam profile when altered by the nonlinear defocusing lens in the VECSEL chip. Such an intensity-dependent lens can lead to beam narrowing at the end mirror of the cavity and favor mode-locking when a slit inserted there adds losses to the cw beam by truncating it laterally. (b) Chip structure with standing-wave electric field at the design wavelength of 960 nm. (c) Band-gap configuration of the different materials used in the gain chip compared to the stopband of the chip and the investigated wavelength range (930 – 975 nm). (d) Measured reflectivity spectrum, surface and edge photoluminescence (PL), and calculated longitudinal confinement factor (LCF) of the investigated VECSEL chip.

## 2. VECSEL chip design and Z-scan measurements

The investigated VECSEL chip sketched in *Figure 1b* consists of 20 (InGa)As/Ga(AsP) quantum wells embedded within GaAs spacer layers to provide partial overlap of the antinodes of the standing wave electric field in the microcavity with the quantum wells. Subsequent 30 GaAs/AlAs distributed Bragg reflector (DBR) pairs provide the high reflectivity needed for efficient laser operation together with external mirrors. The chip is designed for lasing operation around 960 nm. *Figure 1c* shows the band gap alignment of the various materials composing the gain chip. It can be seen that the probe wavelengths, which range from 930 nm to 975 nm, are scanning across the quantum-well band gap, thus, our measurements are particularly sensitive to the nonlinear refractive index changes which are characteristic for near-resonance probing. The wavelength-dependent longitudinal confinement factor (LCF) characterizes the strength of a microcavity's standing-wave electric field at the location of the quantum wells (see Methods) and is depicted together with the reflectivity spectrum, the surface as well as the edge photoluminescence (PL) in *Figure 1d*.

For Z-scan data acquisition and modeling, we refer to the Methods section. The experimental setup is schematically shown in the Supplementary Information together with exemplary measurements. The angle of incidence of the probe beam on the VECSEL chip was varied in the course of the experiment from 10 to 20 and 30°, in order to reveal the influence of the microcavity on the measured nonlinearity. At each angle of incidence, Z-scan measurements were performed for different probe center wavelengths from 930 nm to 975 nm and varying probe intensites.

## 3. Experimental results and discussion

In order to determine very accurately the value of nonlinear absorption and refraction in the VECSEL chip, we performed Z-scan measurements for different probe-peak intensities at center wavelengths from 930 to 975 nm. If a refractive (or an absorbing) third-order nonlinearity is present, the nonlinear phase shift $\Delta\Phi$ (or the normalized nonlinear absorption $q_0$) will vary linearly with peak probe power density. In such case, the strength of the nonlinearity is proportional to the slope of the nonlinear phase shift (or the normalized nonlinear absorption) with respect to the peak probe power density, as presented in the Supplementary Information. Using the relations $\beta = \frac{dq_0}{dI_0}\frac{1}{L_{eff}}$ and $n_2 = \frac{d\Delta\Phi_0}{dI_0}\frac{1}{k\,L_{eff}}$, values for the nonlinear absorption $\beta$ and refraction $n_2$ can be extracted. These are plotted in *Figure 2* for the incidence angles of 10, 20 and 30°.

For both nonlinear absorption and nonlinear refraction, a strong enhancement of the nonlinearity can be observed at wavelengths between 950 to 960 nm. Strikingly, this corresponds very well to the field enhancement at the QWs due to the microcavity resonance, which is represented by the surface photoluminescence peak at that on-chip angle and is also plotted into *Figure 2a and b* for comparison. When increasing the angle of incidence from 10 to 20 and further to 30°, the peak of the nonlinearity spectrum follows the corresponding surface PL spectrum peak, which arises from the longitudinal confinement factor in the microcavity, i.e. the Fabry-Pérot cavity's filter function. This unambiguously demonstrates that the microcavity resonance is strongly contributing to the enhancement of the respective nonlinearity. This enhancement takes place slightly below the linear absorption band edge, which can be estimated from the reflectivity spectra plotted in *Figure 2* for comparison. We consider the contribution of the GaAs spacer layer negligible with respect to the contributions of the (InGa)As/Ga(AsP) gain region, as the theoretical nonlinear refractive index of GaAs[27] is in the order of $-6\cdot10^{-13}$ cm$^2$/W in the investigated wavelength range and thus significantly smaller than the effective $n_2$ values measured here.

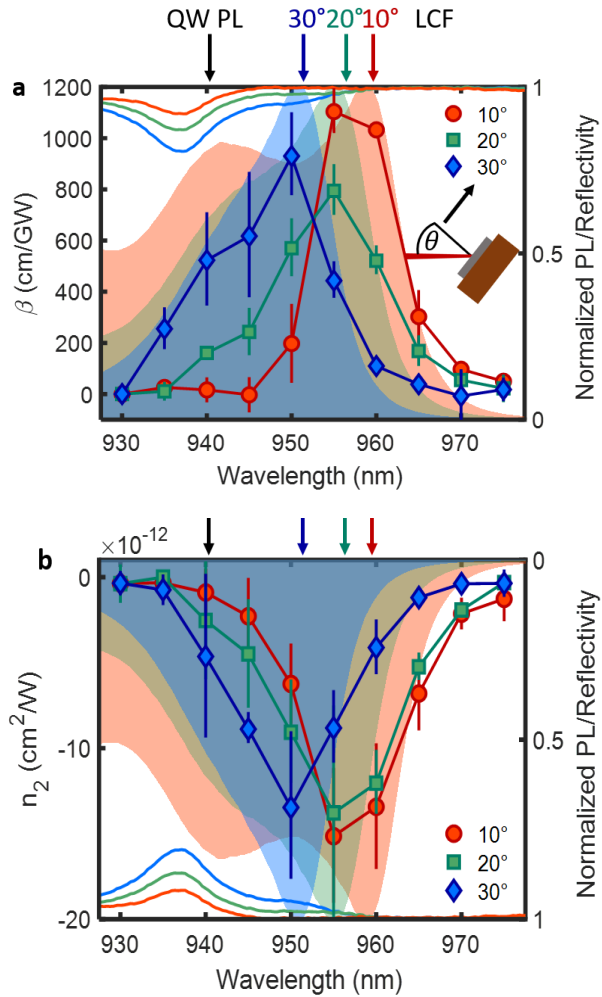

**Figure 2 | Wavelength and angle-dependent nonlinear absorption and nonlinear refraction:** (a) Nonlinear absorption $\beta$ and (b) nonlinear refraction $n_2$ as a function of the wavelength (left axes), measured for the incidence angles of 10, 20 and 30°. The lines between the measurement points may serve as guides to the eyes. Arrows atop the diagram indicate the wavelength of the QW PL (cf. Figure 1d) as well as the different angle-dependent LCF peaks (cf. Figure S.1a). The corresponding surface PL at given angle of incidence, which is subdue to the microcavity resonance, is plotted in both graphs for comparison (shaded plots) as well as the corresponding reflectivity spectra (line plots), both normalized to 1 and with respect to the right axes. An inset represents the probe geometry with respect to the VECSEL chip. The errors were determined by performing a linear least square fit to the normalized nonlinear absorption $q_0$ and nonlinear phase shift $\Delta\Phi$ (as displayed in Figure S.3) and taking the 95% confidence interval of the fit as error.

The absolute magnitude of the effective nonlinear refractive index rises up to $1.5 \cdot 10^{-11}$ cm$^2$/W, which is more than an order of magnitude larger than reported in Ref. [26]. This can be attributed to the fact that the authors did not use a probe wavelength close to the maximum of the microcavity resonance. Moreover, the chip exhibited nearly half the number of quantum wells compared to our chip design.

The strong tunability of the nonlinearity with the microcavity resonance indicates that the (InGa)As/Ga(AsP) material system, which showed best self-mode-locking behavior so far[19], can be possibly operated in the self-mode-locking regime over a large wavelength range exceeding 10 nm thus supporting both broad-band operation and tunability. This is consistent with the results of Ref. [33] for a multiple-quantum-well structure which showed a comparable bandwidth of strong nonlinear refraction (up to $8.5 \cdot 10^{-10}$ cm$^2$/W in magnitude) where no microcavity effect was present.

In order to evaluate our measurement results with respect to possible self-mode-locking in VECSELs due to the Kerr effect, we plot the ratio of real and imaginary part of the third-order nonlinearity $\chi^{(3)}$,

$$\frac{Re(\chi^{(3)})}{Im(\chi^{(3)})} = \left|\frac{2\Delta\Phi_0}{q_0}\right| = \left|\frac{2kn_2}{\beta}\right|, \qquad (4)$$

in *Figure 3* (left axis). This describes how strong nonlinear refraction dominates over nonlinear absorption[8]. We only include wavelengths above 945 nm to cover mostly the laser-relevant wavelength range.

Here, a nearly constant ratio of real and imaginary part of $\chi^{(3)}$ shows that both $n_2$ and $\beta$ are equally affected by the cavity-resonance enhancement. In contrast to this, in pure quantum-well structures without any cavity, this ratio varies considerably as a function of the wavelength around the bandgap[33]. However, optical loss by two-photon absorption is quite significant in our measurements when rising up to more than 15% at the highest probe intensity in *Supplementary Figure S.2c*. Nevertheless, this effect might be reduced in a pumped system, in which population inversion is maintained by the typical above-band high-power continuous-wave pumping scheme. Also, longer pulses – typical pulse lengths of self-mode-locked VECSELs are around 1 ps[21,22] – might reduce the occurrence of two-photon absorption[35].

The typical focal length of a Kerr lens, which would occur in a self-mode-locked VECSEL cavity as reported in Ref. [21], with the here measured value of nonlinear refraction is plotted in the same diagram (right axis of *Figure 3*). For sub-ps pulses with 1-kW peak power, the focal length of the induced Kerr nonlinearity should be in the order of magnitude of the laser cavity length, which ranges usually between 10 to 30 cm. Our calculated values are between a few cm up to 30 cm and more. This information supports the attribution of the observed self-mode-locking being Kerr-lens mode-locking and is key to designing a cavity where the angle of incidence of the laser mode on the gain chip is, apart from geometric constraints, a free design parameter in a V- or Z-cavity.

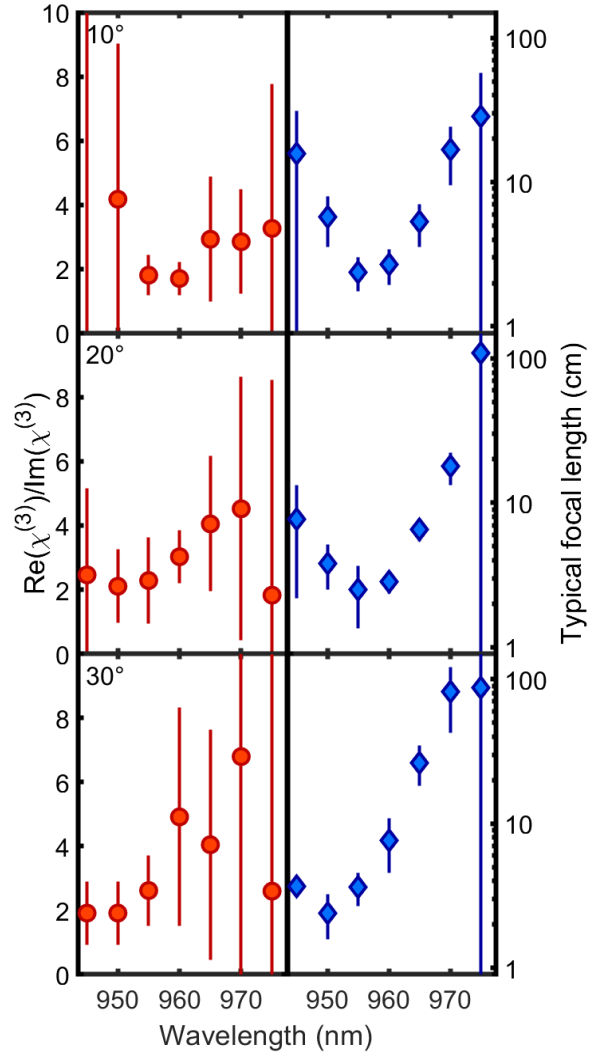

**Figure 3 | Ratio of real and imaginary part of the measured third order nonlinearity $\chi^{(3)}$ and typical focal length of a Kerr lens:** The ratio of nonlinear refraction with respect to nonlinear absorption is plotted as a function of the wavelength in the left column of the diagram for the three different angles of incidence. Additionally, the focal length, which would occur in a typically self-mode-locked VECSEL configuration, is plotted here in the right column. The error bars stem from the errors in nonlinear absorption and refraction as depicted in Figure 2.

### 4. On the origin of nonlinear lensing and the implications for mode-locking

The respective contributions of FCN and BEKE to the total, measured nonlinearity will define the pulse-shaping mechanism in a Kerr-lens mode-locking scenario due to the different response times. FCN has a slower response time, which is mostly due to the ps relaxation time of excited carriers, compared to BEKE, which can be regarded as instantenous with respect to fs pulses. Both FCN and BEKE exhibit characteristic dispersions when regarded separately as depicted in **Figure 4a and c**, with both effects showing strongest changes close to the band gap.

In the theory described by Ref. [27], nonlinear refraction is inherently related to nonlinear absorption by the nonlinear Kramers-Kronig relations. Although in most resonant excitation-and-probe scenarios, carrier-related nonlinearities are believed to dominate the third-order nonlinear response[36], bound-electronic contributions even continue to play a role closely above the band gap when no significant excitation or deexcitation of carriers occur[37]. Strong deexcitation of free carriers could be described with an additional use of the Bloch equations. This has been done in Ref. [38] and showed that a probe pulse will indeed experience a significant phase shift both from FCN and BEKE with the same order of magnitude when the semiconductor material is excited with a high-intensity pump pulse. These findings have also been supported by comparison to experiments[39].

This indicates that also in mode-locked VECSELs a non-negligible amount of BEKE might be present. The occurence of two-photon absorption in our measurements, also above the linear absorption edge, further supports this argument as it indicates that probe-power densities are sufficiently high to significantly perturb pulse propagation by ultrafast $\chi^{(3)}$ effects, both from its imaginary part resulting in two-photon absorption as well as its real part resulting in nonlinear refraction, which as mentioned already above, are linked by nonlinear Kramers-Krönig relations. It is worth noting that also the authors of Ref. [17] attribute their self-mode-locking results with a VECSEL to the ultrafast Kerr effect of bound electrons, without having performed any experimental characterization of the nonlinearity.

However, our measurements did not investigate the effect of cw-pumping, which takes place in actual laser operation. The resulting generation of free carriers, which are probed by the laser pulse and lead to strong stimulated deexcitation of carriers might change the magnitude of the nonlinearity as well as the relative contributions from FCN and BEKE. Preliminary investigations concerning optical pumping and probe-induced nonlinear-refractive-index changes suggest little influence in the absolute value of the nonlinear refraction[26]. The effect of the microcavity resonance on the nonlinearity enhancement revealed by this work will persist in a pumped system.

To ultimately distinguish the possible two contributions from each other, that are the ultrafast bound-electronic Kerr effect (i.e. BEKE) and the non-instantaneous free-carrier nonlinearities (i.e. FCN), time-resolved Z-scan measurements would be required[40]. In addition, the variation of the probe-pulse duration while maintaining equivalent peak powers could provide insight into the nature of the carrier dynamics including the effect of non-equilibrium many-body effects, which become important at sub-ps time-scales[41]. This will be the subject of our future work.

If BEKE is the dominating contribution to the nonlinear lensing, self-mode-locking of VECSELs has to be modeled by substituting the slow, highly nonlinear saturable absorber term with a fast saturable absorber term in the common mode-locking model[42] (c.f. **Figure 4b** on pulse shaping). Additionally, self-phase modulation would have to be incorporated into the model, for which the value of the effective $n_2$ would be obtained from Z-scan measurements. As a consequence of this, the required group-delay dispersion (GDD) for shortest pulses and stable Kerr-lens mode-locking might differ from the slightly positive GDD values required for optimal SESAM-VECSEL mode-locking, which could become a key advantage of self-mode-locked devices towards ultra-short pulse generation after more detailed investigations triggered by the findings reported here.

In contrast, FCN might act as a slow saturable absorber with timescales similar to SESAMs and correspondingly similar mode-locking characteristics. It is important to point out that this would not be due to real saturable absorption occurring in the gain chip. Instead, the refractive-index change induced by changes in the excited-carrier densities would lead to nonlinear lensing and only lead to some kind of artificial saturable-absorber action (c.f. **Figure 4b**) when employed in an appropriate cavity, where nonlinear lensing reduces cavity losses, as indicated in **Figure 1a**.

If both effects were present simultaneously with comparable strength, a combined action of slow and fast saturable absorber would have to be incorporated into a mode-locking model. Also, a situation as in soliton mode-locking might occur in this case[3].

A precise knowledge of the different timescales involved in the effective third-order refractive nonlinearity $n_2$ will ultimately enable us to explain and model the mechanisms behind Kerr-lens induced self-mode-locking, with important implications for future chip designs.

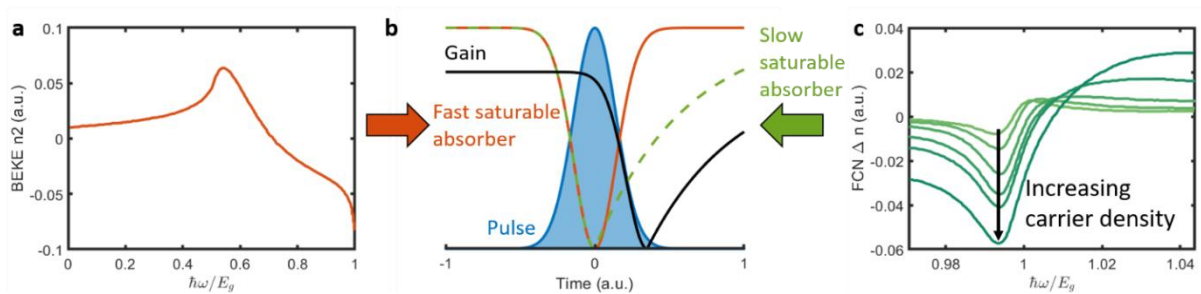

**Figure 4 | Illustration of possible saturable absorber mechanisms shaping the pulse:** A dominating bound-electronic contribution (BEKE) in the nonlinear response, with its dispersion given in (a) as adapted from Ref. [27], would lead to a fast saturable absorber mechanism (red solid line in the schematic time trace of action shown in (b)) similar to traditional Kerr-lens mode-locking. In contrast, a dominating free-carrier-related nonlinear response (FCN), with its dispersion around the band gap given in (c) for different carrier densities as adapted from Ref. [28], would lead to slow saturable absorber action (green dashed line in (b)) where pulse formation relies on its combination with fast gain saturation (black solid line in (b)).

## 5. Conclusions

We have presented unique, systematic and direct measurements of the effective nonlinear refractive index change and nonlinear absorption in a vertical-external-cavity surface-emitting-laser chip using the Z-scan technique. In support of recent self-mode-locking achievements, strong enhancement of the Kerr nonlinearity is observed in the spectral region of the quantum-well band gap in clear correlation to a microcavity effect, which can shape Kerr lensing in the active region significantly. Here, wavelength-tuning of the nonlinearity with the on-chip angle in accordance to the shift of the microcavity resonance is demonstrated. Negative nonlinear refraction up to $1.5 \cdot 10^{-11}$ cm$^2$/W close to the laser design wavelength is obtained. While two-photon absorption poses a loss channel for lasing operation, the refractive index changes are sufficiently strong to perturb the cavity beam profile in favour of self-mode-locking. Moreover, by comparing our results to existing theories about the nonlinear-refractive-index changes in semiconductors, we believe that the ultrafast electronic Kerr effect contributes significantly to the nonlinearity while the share of free-carrier nonlinearities compared to the amount of the ultrafast Kerr effect yet remains to be investigated for a pumped system. We expect our results to lead to a better understanding of self-mode-locking in VECSELs and to pave the way to reliable, ultra-short, higher-power pulsed semiconductor-laser sources based on this method.

## Methods

*Fabrication of the gain chip:*

The semiconductor structure used was fabricated by metal-organic vapor-phase epitaxy (MOVPE) in an AIXTRON FT 3x2" closed coupled shower-head reactor using the standard sources trimethyl-gallium, trimethyl-indium, trimethyl-aluminum, arsine and phosphine. Deposition occured at pressures of 100 mbar on 6° misoriented GaAs substrates. Following a 30 pair AlAs/GaAs DBR the GaAs spacer layer as well as the active region are grown and the whole structure is completed using a GaAs capping layer.

The active region comprises 20 (InGa)As quantum wells grouped into five packages which are separated using (AlGa)As spacer layers. The compressive strain induced by the QWs is mostly compensated by the surrounding tensile-strained Ga(AsP) barriers and the QW packages are placed according to a resonant periodic gain design at the antinodes of the standing wave electric field.

*E-field simulations and longitudional confinement factor:*

To calculate the electric-field distribution in the structure for a particular wavelength, we use the matrix formalism of Ref. [43]. The overlap of the electric field with the position of the quantum wells is then described by the longitudinal confinement factor, $\Gamma_z = \frac{\sum_q |A_{i_q}^+ \exp(ik_{i_q}z_q) + A_{i_q}^- \exp(-ik_{i_q}z_q)|^2}{|A_0^+|^2 + |A_0^-|^2}$, where $A_{i_q}^+$ and $A_{i_q}^-$ are the forward and backward traveling wave amplitudes of the vector potential of the electric field at the q[th] quantum well in the i[th] material layer, respectively. $k_{i_q}$ is the respective wave vector and $z_q$ the position of the q[th] QW. $A_0^+$ and $A_0^-$ are the vector potential amplitudes before the first material layer. The matrix formalism is solved for TE-polarization in accordance with the polarization of the probe laser.

*Z-scan setup:*

For Z-scan measurements, the chip structure was mounted on a piece of copper without any active cooling. A wavelength tunable, mode-locked Ti:Sapphire laser emitting pulses of approximately 150 fs pulse length and a spectral width of about 8 nm was used to probe the chip structure at different center wavelengths around the design wavelength of the gain chip.

For each center wavelength, the pulse length was determined with a frequency-resolved optical gating (FROG) device and the laser spectrum was monitored with an optical spectrometer.

A lens with 50 mm focal length was used to focus the probe beam down to a spot size of 12 µm which was confirmed by a beam profile measurement around the focus of the probe beam. After reflection from the sample and a folding mirror, which were both mounted on a translation stage, a beam splitter and two detectors were used to record the position-dependent transmittance of the probe laser when the sample was moved through the focus of the lens. One detector was covered by a partially closed aperture, which was aligned for maximal transmission. The second detector collected the whole probe beam using a focusing lens. For further setup details and a schematic drawing of the experiment, we refer to the Supplementary Information.

*Data modeling:*

In order to extract the nonlinear absorption coefficient $\beta$, we fit the open scan data with the model from Ref. 8,

$$T(z) = \sum_{m=0}^{\infty} \frac{[-q_0(z)]^m}{(m+1)^{\frac{3}{2}}}, \quad (2)$$

with $q_0(z) = \beta I(z) L_{eff} = \frac{\beta I_0 L_{eff}}{1+\frac{z^2}{z_0^2}}$ being the normalized nonlinear absorption and $I(z) = \frac{I_0}{1+\frac{z^2}{z_0^2}}$ the z-dependent peak-probe power density as a function of z and $I_0$, $L_{eff}$ and $z_0$ being the peak probe power density, the effective length of the sample and the Rayleigh length of the focused probe beam, respectively. The effective length $L_{eff} = \frac{1-e^{-\alpha_0 L}}{L}$ takes into account the different strength of linear absorption $\alpha_0$ at different wavelengths. The total length $L$ is here the accumulated thickness of the different material layers including the penetration depth into the DBR and taking into account the double-pass through the chip with the respective angle of incidence. This leads to lengths $L$ from 4.4 to 5.1 µm for angles of incidence from 10 to 30°. Saturable absorption, which becomes important at wavelengths below 950 nm, is taken into account with an additional term as detailed in the supplement.

We extract the nonlinear phase shift $\Delta\Phi$ from the Z-scans by fitting the normalized transmission to the simple model from Ref. 8,

$$T(z) = 1 - \frac{4\Delta\Phi \frac{z}{z_0}}{\left(\left(\frac{z}{z_0}\right)^2 + 9\right)\left(\left(\frac{z}{z_0}\right)^2 + 1\right)}, \qquad (3)$$

where $\Delta\Phi = k n_2 I_0 L_{eff}$ is the nonlinear phase shift induced by the nonlinear refractive index $n_2$. The sign of the nonlinear phase shift can be easily determined by the position of peak and valley in the Z-scan trace around the focus, which in our measurements always leads to a negative, thus defocusing, nonlinearity.

**Acknowledgement**

The authors thank S.W. Koch and M. Gaafar for helpful discussions and O. Mohiuddin as well as M. Alvi for technical assistance. This work was funded by the Deutsche Forschungsgemeinschaft (DFG) under Grant No. RA2841/1-1.

**Authors' contributions**

A.R.-I. initiated the study and conceived a project on self-mode-locking in VECSELs and investigations of its underlying mechanisms in 2015. Experimental work was performed by C.K., S.K. and T.M. under the guidance of M.K. and A.R.-I. The measurement setup was established by S.K. using a VECSEL structure from W.S. and the system was expanded by C.K. and T.M with support by S.K., M.K. and A.R.-I. The data was analyzed by C.K. and the results were interpreted by C.K. and A.R.-I. with the help of all coauthors.  R.B., M.J. and P.M. designed and grew the test sample, M.G. contributed simulations on the chip's electric-field distribution. The manuscript was written by C.K. and A.R.-I with the help of all coauthors.

**Corresponding author**

a.r-i@physik.uni-marburg.de

**Authors' statement/Competing interests**

The authors declare no conflict of interest

**Additional information**

Supplementary Information accompanies this paper

*Figure Captions*

**Figure 1 | VECSEL-chip characteristics:** (a) Illustration of a Kerr-lens mode-locked VECSEL with the dashed lines indicating the cw beam profile while the shaded area traces the beam profile when altered by the nonlinear defocusing lens in the VECSEL chip. Such an intensity-dependent lens can lead to beam narrowing at the end mirror of the cavity and favor mode-locking when a slit inserted there adds losses to the cw beam by truncating it laterally. (b) Chip structure with standing-wave electric field at the design wavelength of 960 nm. (c) Band-gap configuration of the different materials used in the gain chip compared to the stopband of the chip and the investigated wavelength range (930 – 975 nm). (d) Measured reflectivity spectrum, surface and edge photoluminescence (PL), and calculated longitudional confinement factor (LCF) of the investigated VECSEL chip.

**Figure 2 | Wavelength and angle-dependent nonlinear absorption and nonlinear refraction:** (a) Nonlinear absorption $\beta$ and (b) nonlinear refraction $n_2$ as a function of the wavelength (left axes), measured for the incidence angles of 10, 20 and 30°. The lines between the measurement points may serve as guides to the eyes. Arrows atop the diagram indicate the wavelength of the QW PL (cf. Figure 1d) as well as the different angle-dependent LCF peaks (cf. Figure S.1a). The corresponding surface PL at given angle of incidence, which is subdue to the microcavity resonance, is plotted in both graphs for comparison (shaded plots) as well as the corresponding reflectivity spectra (line plots), both normalized to 1 and with respect to the right axes. An inset represents the probe geometry with respect to the VECSEL chip. The errors were determined by performing a linear least square fit to the normalized nonlinear absorption $q_0$ and nonlinear phase shift $\Delta\Phi$ (as displayed in Figure S.3) and taking the 95% confidence interval of the fit as error.

**Figure 3 | Ratio of real and imaginary part of the measured third order nonlinearity $\chi^{(3)}$ and typical focal length of a Kerr lens:** The ratio of nonlinear refraction with respect to nonlinear absorption is plotted as a function of the wavelength in the left column of the diagram for the three different angles of incidence. Additionally, the focal length, which would occur in a typically self-mode-locked VECSEL configuration, is plotted here in the right column. The error bars stem from the errors in nonlinear absorption and refraction as depicted in Figure 2.

**Figure 4 | Illustration of possible saturable absorber mechanisms shaping the pulse:** A dominating bound-electronic contribution (BEKE) in the nonlinear response, with its dispersion given in (a) as adapted from Ref. [27], would lead to a fast saturable absorber mechanism (red solid line in the schematic time trace of action shown in (b)) similar to traditional Kerr-lens mode-locking. In contrast, a dominating free-carrier-related nonlinear response (FCN), with its dispersion around the band gap given in (c) for different carrier densities as adapted from Ref. [28], would lead to slow saturable absorber action (green dashed line in (b)) where pulse formation relies on its combination with fast gain saturation (black solid line in (b)).

# Supplementary Information

**Microcavity-enhanced Kerr nonlinearity in a vertical-external-cavity surface-emitting laser**

C. Kriso[1], S. Kress[1], T. Munshi[1], M. Großmann[2], R. Bek[2], M. Jetter[2], P. Michler[2], W. Stolz[1], M. Koch[1], and A. Rahimi-Iman[1]

[1]*Faculty of Physics and Materials Sciences Center, Philipps-Universität Marburg, D-35032 Marburg, Germany*
[2]*Institut für Halbleiteroptik und Funktionelle Grenzflächen, Universität Stuttgart, D-70569 Stuttgart, Germany*

**S1: Angle-dependent VECSEL-chip properties**

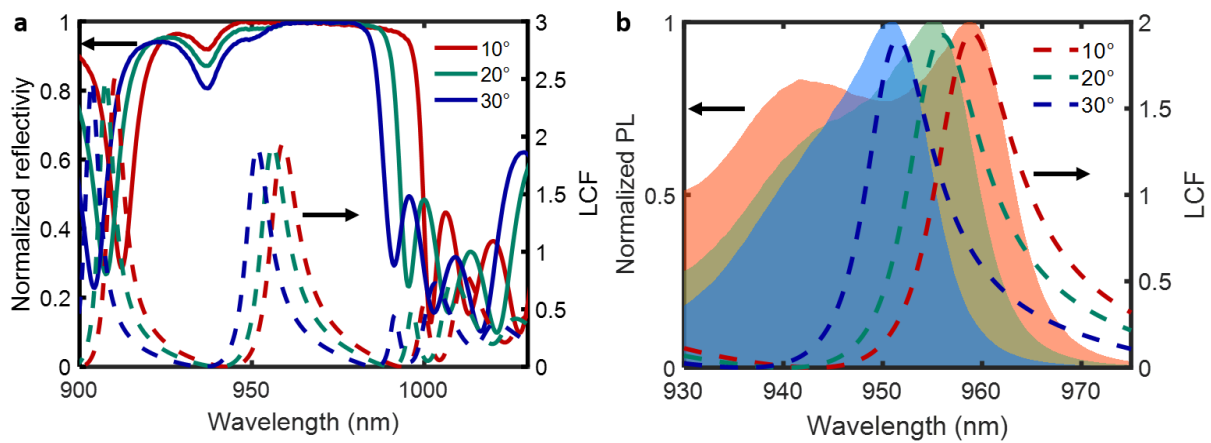

**Figure S.1| Angle-dependent chip characterization:** (a) Measured reflectivity spectra for different angles of incidence (solid lines, left axis) as well as calculated longitudinal confinement factor (LCF) for the same angles (dashed lines, right axis). (b) Zoom on the longitudinal confinement factor (dashed lines) around the relevant wavelength range in comparison to the measured surface photoluminescence (PL) at the same angles (shaded areas).

The angle-dependent characteristics of the Bragg reflector and the resonant-periodic-gain structure are displayed in *Figure S.1*. *Figure S.1a* shows the full stopband of the DBR for different angles of incidence together with the corresponding calculated longitudinal confinement factors. For an increasing angle of incidence, both the stopband and the cavity resonance experience a significant blue shift.

The surface PL spectra in the *Figure S.1b* (the same as in *Figure 2a and b*) are recorded by placing the collection optics (collimating and focusing lens aligned in a lens tube) at the respective angle of incidence with respect to the gain chip and by fiber coupling the signal to an optical spectrum analyzer. These spectra correspond well with the respective calculated longitudinal confinement factors, thus, validating the use of the surface PL for determining the spectral position of the microcavity resonance.

**S2: Experimental setup and exemplary data**

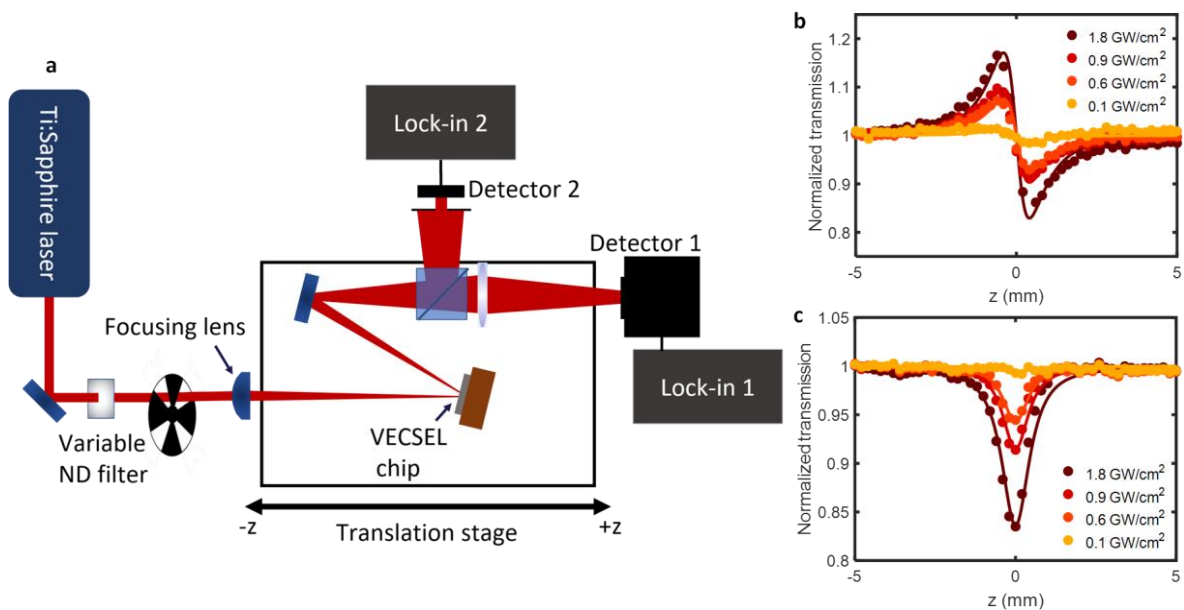

**Figure S.2 | Experimental Z-scan setup and data**: Different neutral-density (ND) filters are used to systematically vary the incident probe power on the VECSEL chip. The sample is translated through the focus of the 150-fs pulsed probe beam, which is chopped and analyzed by two detectors using a lock-in scheme. (b) Exemplary Z-scan measurements at 955 nm for different peak probe intensities irradiated at an angle of 20° are shown, with the corresponding open scan recorded on detector 1 (c). The solid lines are the respective fits to the measurement data according to the models from Ref. [8].

The experimental setup is schematically shown in *Figure S.2a*. The sample moves through the focus of the mode-locked laser beam which can be attenuated by introducing neutral-density filters. After reflection from the VECSEL chip, detector 1 collects the total beam cross section and thus accounts only for nonlinear absorption. The transmittance as a function of the sample position will be referred to as open scan in the following. Detector 2, partially covered by an aperture, is sensitive to, both, nonlinear lensing and nonlinear absorption. The resulting trace will be referred to as closed scan. Following Ref. 8, we obtain the final Z-scan curve by dividing the closed scan by the open scan, performing a background subtraction with a low-intensity scan and normalizing to a transmittance of 1 for transmission at positions far away from the focus.

*Figure S.2b* shows exemplary Z-scan measurements at a center wavelength of 955 nm for different peak probe intensities at 20° angle of incidence together with the respective fits. They show good agreement between model and experiment including the symmetric decrease of peak and valley transmission for decreasing peak probe-power density. The decrease in transmission close to the focus in the open scan (*Figure S.2c*) is attributed to two-photon absorption.

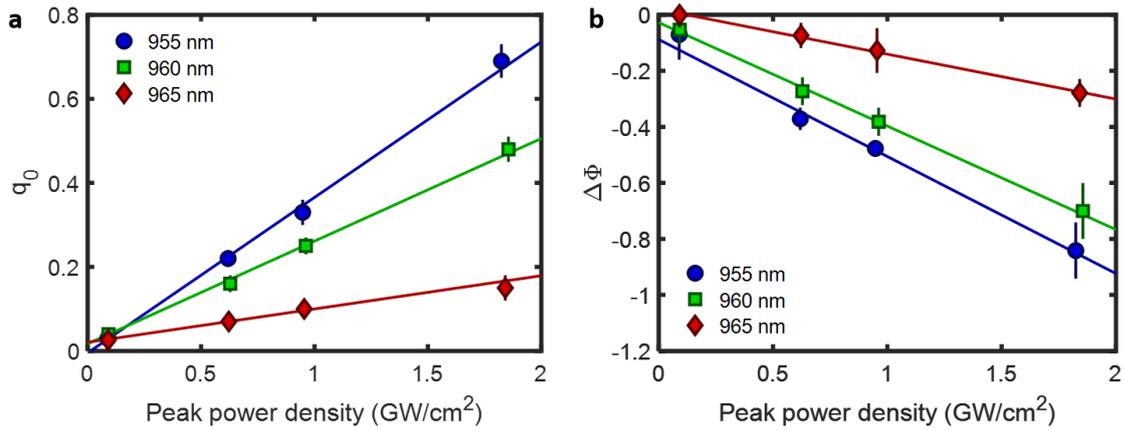

**Figure S.3 | Coefficients for the extraction of nonlinear absorption and refraction:** (a) Normalized nonlinear absorption $q_0$ and (b) nonlinear phase shift $\Delta\Phi$ as a function of the incident peak probe intensity for different probe center wavelengths irradiated at 20°. The errors for the respective extracted magnitudes represent the fit tolerance and were determined by varying the fit parameter ($q_0$ or $\Delta\Phi$) until the fit model did not match the measurements anymore (similar to Ref. [44]).

*Figure S.3* shows, both, the normalized nonlinear absorption $q_0$ (*Figure S.3a*) and the nonlinear phase shift $\Delta\Phi$ (*Figure S.3b*) as a function of the incident peak power density. The very good agreement with the linear fit shows that measured nonlinear absorption and nonlinear phase shift are indeed mainly caused by a third-order nonlinearity. It is important to note, that nonlinear-refractive-index changes caused by two-photon-absorption-generated free carriers would appear as an effective fifth-order nonlinearity[8] causing a deviation from the linear fit. Due to the good agreement to the linear fit, this mechanism can be neglected for our measurements.

**S3: Simultaneous presence of saturable absorption and two-photon absorption in open scan data**

Saturable absorption, which becomes important above the linear absorption edge at wavelengths below 950 nm is phenomenologically included to the model applied on the open-scan data by adding the simple model for saturable absorption in transmission,

$$T_{sat}(z) = 1 - \frac{\alpha_0}{1+\frac{I(z)}{I_s}}, \tag{S1}$$

to Eq. (2). Here, $\alpha_0$ is again the linear absorption at the respective center wavelength of the probe laser spectrum, and $I_s$ the saturation intensity which is determined by the fit. Eq. (S1) is normalized to 1 at the edges of the scan, added to Eq. (2) and subtracted by 1 to align the total open-scan fit function to a transmission of 1 for z positions far away from the focus. An exemplary fit of an open scan at 940 nm for 20° angle of incidence is displayed in *Figure S.4a*. The resulting normalized nonlinear absorption (the positive one attributed to two-photon absorption) exhibits a good linear trend as shown in *Figure S.4b*, thus validating our method.

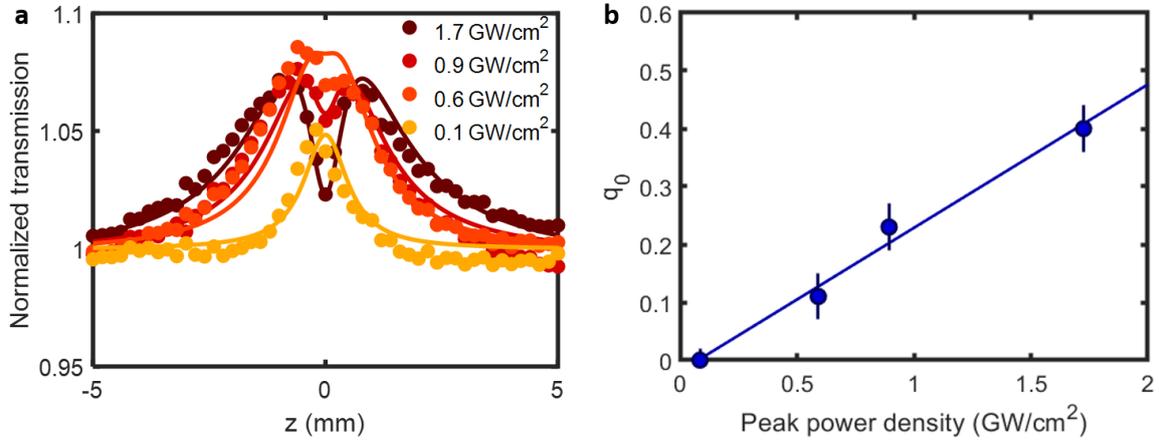

**Figure S.4 | Example of extraction of normalized nonlinear absorption in the presence of saturable absorption:** (a) Open scan at 940 nm and 30° angle of incidence for different probe powers. The fit (solid line) is obtained as described in section 3 using $I_s$=0.2 GW/cm² and $\alpha_0$=1.2e-5 m⁻¹ for all probe intensities. (b) Resulting nonlinear absorption as a function of the peak probe density with linear fit (solid line).

**S4: Determination of fit uncertainty in open scan and Z-scan**

To estimate the uncertainty in the determination of the fit parameter $q_0$ or $\Delta\Phi$, we vary them in Eq. (2) and (3) until there is a significant deviation from the measurement data. Upper and lower bounds are taken as equally spaced from the fit. An example of the limiting cases is displayed in *Figures S.5a and b* for open scan and Z-scan of *Figure S.2*. The bounds of $q_0$ and $\Delta\Phi$ are then taken into account in a weighted linear least square fit where the weights are the inverse of the respective variances meaning they are proportional to the inverse of the square of the error bounds. The resulting fits are plotted in *Figure S.3*. The subsequent error of the value of nonlinear refraction $n_2$ or $\beta$ as displayed in *Figure 2* is then the 95 % confidence interval of this fit.

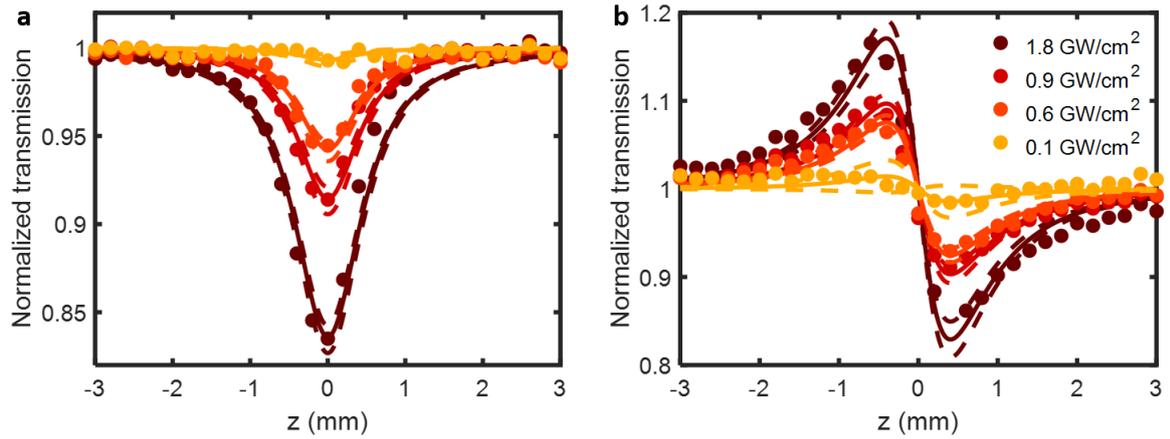

**Figure S.5 | Measurement data with fit model as well as manually-determined symmetrical upper and lower bounds of the fit:** This is a zoom in of Figure S.2b and c. (a) Open scan and (b) Z-scan with fit (solid line) and upper and lower limit of the respective fits (dashed lines) for the estimation of the data-evaluation uncertainties.